\documentclass[12pt]{article}

\newcommand{\be}{\begin{equation}}
\newcommand{\ee}{\end{equation}}

\usepackage{epsfig,amsmath,amssymb,graphics,graphicx}

\usepackage{amscd}
\usepackage{pb-diagram}
\usepackage{hyperref}
\usepackage{amsfonts}

\begin{document}


Physics Letters A. 2020. Article ID: 126303.
DOI: 10.1016/j.physleta.2020.126303

\begin{center}

{\bf \large Dirac Particle with Memory: 
Proper Time Non-Locality} 

\vskip 7mm
{\bf \large Vasily E. Tarasov}$^{1,2}$ \\
\vskip 3mm

${}^1$ {\it Skobeltsyn Institute of Nuclear Physics, \\ 
Lomonosov Moscow State University, Moscow 119991, Russia} \\
{E-mail: tarasov@theory.sinp.msu.ru} \\

${}^2$ {\it Faculty "Information Technologies and Applied Mathematics", \\ Moscow Aviation Institute (National Research University), Moscow 125993, Russia } \\

\begin{abstract}
A generalization of the standard model of Dirac particle in external electromagnetic field is proposed.
In the generalization we take into account interactions of this particle with environment, which is described by the memory function. This function takes into account that the behavior of the particle at proper time can depend not only at the present time, but also on the history of changes on finite time interval.
In this case the Dirac particle can be considered an open quantum system with non-Markovian dynamics.
The violation of the semigroup property of dynamic maps is a characteristic property of dynamics with memory.
We use the Fock-Schwinger proper time method and derivatives of non-integer orders with respect to proper time.
The fractional differential equation, which describes the Dirac particle with memory, 
and the expression of its exact solution are suggested. The asymptotic behavior of the proposed solutions is described.
\end{abstract}

\end{center}



\section{Introduction}

The Dirac particles can be described by the Dirac equation.
This approach corresponds to the Schrodinger picture. 
The Dirac particles can also be described by using the Heisenberg picture.
This approach was proposed by Vladimir A. Fock \cite{PT1,PT2,PT3,PT4,PT4E} and Julian S. Schwinger in \cite{PT5}.
In this case, the differential equations with respect to proper time are used to have the Hamilton and Heisenberg equations in the covariant form \cite{IZ,PT7,PT7b}. This approach is called the Fock-Schwinger proper time method \cite{IZ,IZ2,IZ3}.
In the standard model the Dirac particle is described in the external electromagnetic field (for example, see \cite{PT5},
\cite{IZ}, p.100-104, and \cite{IZ3}, p.728-732). 
The standard theory of the Dirac particles does not take into account an interaction between this particle and the environment. 

We proposed a generalization of the Heisenberg equations for the Dirac particle in the external electromagnetic field.
In this generalization we take into account an interaction of this particle with environment.
The interaction with environment can be described by the memory function \cite{CEJP2012}, that is also called the response function.
Therefore the proposed model can be considered an open quantum system, and the dynamics of this Dirac particle is non-Markovian.
The Dirac particle is considered in the Heisenberg picture.
In this paper we will use the Fock-Schwinger proper time method
to preserve the relativistically covariant form of Heisenberg equations.
The proposed Heisenberg equations are fractional differential equation with derivatives of non-integer orders with respect to proper time.
The exact expressions of the solution of the generalized Heisenberg equations are suggested for 
particular case of the external electromagnetic field (the constant uniform field). 
The standard solution for the absence of memory is the special case of the proposed solution.
In this article also suggests the asymptotic behavior of the solution as the proper time tends to infinity.

From a physical point of view, the neglect of memory in the standard model is based on the assumption that the Dirac particle is not open system. In the standard approach the interaction of Dirac particle with environment is not considered.
The first description of physical system with memory was suggested by Ludwig Boltzmann in 1874 and 1876. He took into account that the stress at time $t$ can depend on the strains not only at the present time $t$, but also on the history of changes at $t_1<t$. Boltzmann also proposed the linear position principle for memory and the memory fading principle. The linear position principle states that total strain is a linear sum of strains, which have arisen in the media at $t_1<t$. The principle of memory fading states that the increasing of the time interval lead to a decrease in the corresponding contribution to the stress at time $t$. The Boltzmann theory has been significantly developed in the works of the Vito Volterra in 1928 and 1930 in the form of the heredity concept and its application to physics. 
Then the memory effects in natural and social sciences are considered in different works (for example, see \cite{M1}-\cite{M9}).
 As example, we can note the thermodynamics of materials with fading memory \cite{M1,M2}, the hereditary solid mechanics \cite{M3}, the theory of wave propagation in media with memory \cite{M4}, the dynamics of media with viscoelasticity \cite{M5}, the econophysics of processes with memory \cite{M6}, 
the condensed matter with memory \cite{M7}, 
the fractional dynamics of systems and processes with memory \cite{M8,M9}.

From a mathematical point of view, the neglect of memory effects is due to the fact that to describe the system we use only equations with derivatives of an integer order, which are determined by the properties of the function in an infinitely small neighborhood of the considered time. 
An effective and powerful tool for describing memory effects is the fractional calculus of derivatives and integrals of non-integer orders \cite{FC1,FC2,FC3,FC4,FC5}. These operators have a long history of more than three hundred years \cite{HFC1,HFC2,HFC3,HFC4,HFC5}.
We should note that the characteristic properties of fractional derivatives of non-integer order are the violation of standard rules and properties that are fulfilled for derivatives of integer order \cite{PFC1,PFC2,PFC3,PFC4,PFC5}. 
For example, the standard product (Leibniz) rule and the standard chain rule are violated for derivatives of non-integer orders.
These non-standard mathematical properties allow us to describe non-standard processes and phenomena associated with non-locality and memory \cite{M8,M9}. 
On the other hand, these non-standard properties lead to difficulties \cite{Rules} in sequential constructing a fractional generalizations of standard models. 
In article \cite{Rules}, we show how problems arise when building fractional generalizations of standard models.

In this paper, we proposed a generalization of the standard model of the Dirac particle in the external electromagnetic field by taking into account the interaction with an environment that is described by memory function.
The fractional nonlinear differential equation, which describes the proposed model, and the expression of its exact solution are suggested.
The asymptotic behavior of the proposed solutions is described.


\section{Dirac particle in external electromagnetic field}

The kinematics of the Dirac particle can be described by the commutation relations 
\begin{equation} \label{DP1}
[x^{\mu},p^{\nu}] = - i g^{\mu \nu} ,
\end{equation}
where $p_{\mu} = i \partial_{\mu}$ with where $\mu=0,1,2,3$, and
\begin{equation} \label{DP2}
<x|x^{\prime}> =\delta^{4} (x-x^{\prime}) .
\end{equation}
 We use the notation of the book \cite{IZ}, 
\begin{equation} \label{DP2g}
g^{00}=-g^{11}=-g^{22}=-g^{33}=1 ,
\end{equation}
and $A^0=A_0$, $A^k=-A_k$.

Let us consider the standard function ${\cal H}(x, p)$ in the coordinate representation 
\begin{equation} \label{DP3}
{\cal H}= (p - e A)^2 - m^2 - 
\frac{e}{2} \sigma_{\mu \nu} F^{\mu \nu} ,
\end{equation}
where $F^{\mu \nu}$ is the antisymmetric field strength tensor
\begin{equation} \label{DP4}
 F^{\mu \nu} = \partial^{\mu} A^{\nu} - \partial^{\nu} A^{\mu} ,
\end{equation}
and
\begin{equation} \label{DP5}
\sigma^{\mu \nu} = \frac{1}{2} i [\gamma^{\mu},\gamma^{\nu}] .
\end{equation}
The electromagnetic field strength tensor $F^{\mu \nu}$
can be expressed through the components of the
electric field ($E^k$) and magnetic field $B^k$, where $k=1,2,3$, in the form
\begin{equation} \label{DP5f} 
F^{\mu \nu} = \left( \begin{array}{cccc}
0 & -E^1 & -E^2 & -E^3 \\ 
E^1 & 0 & -B^3 & B^2 \\ 
E^2 & B^3 & 0 & -B^1 \\ 
E^3 & -B^2 & B^1 & 0 
\end{array}
\right) . 
\end{equation}

The idea of the Fock-Schwinger proper time method 
is to consider ${\cal H}$ as Hamiltonian that describes the proper time evolution of some system.

Using the operator
\begin{equation} \label{DP6}
\pi_{\mu} = p_{\mu} - e A_{\mu} ,
\end{equation}
the standard Heisenberg equations for $x_{\mu} (\tau)$ and $\pi_{\mu} (\tau)$ of the Dirac particle in the external electromagnetic field can be written in the form
\begin{equation} \label{DP7}
\frac{dx_{\mu} (\tau)}{d \tau} = i [{\cal H}, x_{\mu}(\tau)] ,
\end{equation}
\begin{equation} \label{DP8}
\frac{d\pi_{\mu} (\tau)}{d \tau} = 
i [{\cal H}, \pi_{\mu}(\tau)] ,
\end{equation}
where $\tau$ is proper time of the Dirac particle.
Using the Hamiltonian
\begin{equation} \label{DP9}
{\cal H}= \pi^2 - m^2 - 
\frac{e}{2} \sigma_{\mu \nu} F^{\mu \nu} ,
\end{equation}
and the commutation relations
\begin{equation} \label{DP10a}
[\pi^{\mu},\pi^{\nu}] = - i e F^{\mu,\nu} ,
\end{equation}
\begin{equation} \label{DP10b}
[x^{\mu},\pi^{\nu}] = - i g^{\mu \nu} ,
\end{equation}
\begin{equation} \label{DP11}
[x^{\mu},x^{\nu}] = 0 ,
\end{equation}
the Heisenberg equations of the Dirac particle takes the form
\begin{equation} \label{DP12}
\frac{dx_{\mu} (\tau)}{d \tau} = - 2 \pi_{\mu} ,
\end{equation}
\begin{equation} \label{DP13}
\frac{d\pi_{\mu} (\tau)}{d \tau} = - 2 e F_{\mu \nu} \pi^{\nu} 
-ie \partial^{\nu} F_{\mu \nu}
- \frac{e}{2} \partial_{\mu} F_{\nu \rho} \sigma^{\nu \rho} .
\end{equation}
Note that the coefficient "-2" in equations \eqref{DP12} and
\eqref{DP13} is due to the form of the operator \eqref{DP6}
the Hamiltonian \eqref{DP9} that are standard for the theory of Dirac particle in external electromagnetic field 
(see original Schwinger's article \cite{PT5}, 
Section 2.5.4 in \cite{IZ}, p.100-104, and 
Section 33 in \cite{IZ3}, p.728-732).

The dynamics of quantum observables $x_{\mu} (\tau)$ and $\pi_{\mu} (\tau)$ can be described by as unitary evolution in the form
\begin{equation} \label{DP14}
x_{\mu} (\tau) = U^*(\tau) x_{\mu}(0) U(\tau), \quad
\pi_{\mu} (\tau) = U^*(\tau) \pi_{\mu}(0) U(\tau),
\end{equation}
where $U(\tau)$ the unitary operator \cite{IZ}.

Note that in general the evolution cannot be considered as the unitary evolution as in the standard model. 
Dynamics with memory (with proper time non-locality) cannot be considered as the unitary evolution. 
The evolution of fractional dynamic systems with memory cannot be described by unitary operators. 
The dynamics with memory is non-Markovian in general.


\section{Dirac particle with memory in external electromagnetic field}

The first description of physical system with memory was suggested by Ludwig Boltzmann in 1874 and 1876.
He takes into account that the stress at time $t$ can depend on the strains not only at the present time $t$, but also on the history of changes at $t_1<t$. The Boltzmann theory has been significantly developed in the works of the Vito Volterra in 1928 and 1930 in the form of the heredity concept and its application to physics. 
Then the memory effects in natural and social sciences are considered in different works (for example, see \cite{M1}-\cite{M9}).

To take into account the memory effects for Dirac particle in external electromagnetic field, we can consider the equations
\begin{equation} \label{DP15}
{\cal E}^{\tau}_{0;M,n} \{x_{\mu}\} = i [{\cal H}, x_{\mu}(\tau)] = - 2 \pi_{\mu} ,
\end{equation}
\begin{equation} \label{DP16}
{\cal E}^{\tau}_{0;M,n} \{\pi_{\mu}\} = i [{\cal H}, \pi_{\mu}(\tau)] 
 = - 2 eF_{\mu \nu} \pi^{\nu} -ie \partial^{\nu} F_{\mu \nu}
- \frac{e}{2} \partial_{\mu} F_{\nu \rho} \sigma^{\nu \rho} ,
\end{equation}
where the symbol ${\cal E}^{\tau}_{0;M,n}$ denotes a certain operator that allows us to find the values of $(x_{\mu}(\tau),\pi_{\mu}(\tau))$ for any proper time $\tau$, if the values $(x_{\mu}(\tau_1),\pi_{\mu}(\tau_1))$ are known for $\tau_1 \in [0,\tau]$. We can say that ${\cal E}^{\tau}_{0;M,n}$ is an operator, which is a mapping from one space of functions to another. 
In this paper, we consider linear operators ${\cal E}^{\tau}_{0;M,n}$ of a special kind, which is called the Volterra operator that is defined by the expression
\begin{equation} \label{DP17}
{\cal E}^{\tau}_{0;M,n} \{{\cal A}_{\mu}\} = \int^{\tau}_0 M(\tau - \tau_1) {\cal A}^{(n)}_{\mu} (\tau_1) d \tau_1 ,
\end{equation}
where ${\cal A}^{(n)}_{\mu} (\tau)= d^n {\cal A}_{\mu} (\tau) / d\tau^n$ is the derivative of 
the positive integer order $n \in \mathbb{N}$ with respect to proper time.
The function $M(\tau - \tau_1)$, which characterizes an interaction of the particle with environment, is called the memory function. 
In statistical mechanics it is also called the response function.

It is obvious that not every operators ${\cal E}^{\tau}_{0;M,n}$ of the form \eqref{DP17} can be used to describe a memory. 
Restrictions on the memory functions and consideration of the examples of memory function are discussed in paper \cite{Concept}.

In paper \cite{Concept}, some general restrictions that can be imposed on the properties of memory are described. 
In addition, to the causality principle, these restrictions include the following three principles: the principle of memory fading; the principle of memory homogeneity on time (the principle of non-aging memory); the principle of memory reversibility (the principle of memory recovery).

For example, in the operator \eqref{DP17} 
we can consider the memory function of the power-law form
\begin{equation} \label{DP18}
M (\tau - \tau_1) =
\frac{1}{\Gamma (n-\alpha)} (\tau - \tau_1)^{n-\alpha-1} ,
\end{equation}
where $\alpha>0$.
In the case of integer values of $\alpha$, the operator ${\cal E}^{\tau}_{0;M,n} \{{\cal A}_{\mu}\}$ is the derivative of integer order
\begin{equation} \label{DP19}
{\cal E}^{\tau}_{0;M,n} \{{\cal A}_{\mu}\} = 
\frac{d^n {\cal A}_{\mu} (\tau) }{d \tau^n} .
\end{equation}

In the case of positive real values of $\alpha \in \mathbb{R}_+$, the operator ${\cal E}^{\tau}_{0;M,n} \{{\cal A}_{\mu}\}$ is the Caputo fractional derivative of the order $\alpha>0$. 
The Caputo fractional derivative \cite{FC4} is defined by the equation
\begin{equation} \label{DP22} 
\left({\cal D}^{\alpha}_{\tau;0+} {\cal A}_{\mu}\right)(\tau) = 
\frac{1}{\Gamma (n-\alpha)}
\int^{\tau}_{0} (\tau - \tau_1)^{n-\alpha -1} {\cal A}^{(n)}(\tau_1) d\tau_1 , 
\end{equation} 
where $n=[\alpha]+1$ for non-integer values of $\alpha$, and $n =\alpha$ for $\alpha \in \mathbb{N}$, (see equation 2.4.3 in \cite{FC4}, p.91), $\Gamma (\alpha)$ is the Gamma function, and 
${\cal A}^{(n)}_{\mu} (\tau)$ is the derivative of the integer order $n$. It is assumed that the function 
${\cal A}_{\mu}(\tau)$ has derivatives up to the $(n-1)$th order, 
which are absolutely continuous functions on a finite time interval \cite{FC4}.
If $\alpha=n \in \mathbb{N}$ and the usual derivative ${\cal A}^{(n)}_{\mu} (\tau)$ of order $n$ exists, then the Caputo fractional derivative \eqref{DP22} coincides with ${\cal A}^{(n)}_{\mu} (\tau)$ (for example, see equation 2.4.14 in \cite{FC4}, p.92, proof in \cite{FC3}, p.79), i.e.
the condition \eqref{DP19} holds.

For the power-law memory function \eqref{DP17}, equations \eqref{DP15} and \eqref{DP16} take the form
\begin{equation} \label{DP20}
\left({\cal D}^{\alpha}_{\tau,0+} x_{\mu} \right) (\tau) = 
i [{\cal H}, x_{\mu}(\tau)] = - 2 \pi_{\mu} ,
\end{equation}
\begin{equation} \label{DP21}
\left({\cal D}^{\alpha}_{\tau,0+} \pi_{\mu} \right) (\tau) 
= i [{\cal H}, \pi_{\mu}(\tau)] 
 = - 2 eF_{\mu \nu} \pi^{\nu} -ie \partial^{\nu} F_{\mu \nu}
- \frac{e}{2} \partial_{\mu} F_{\nu \rho} \sigma^{\nu \rho} ,
\end{equation}
where ${\cal D}^{\alpha}_{\tau,0+}$ is the Caputo fractional derivative of the order $\alpha>0$.
In the general case, we can consider the different orders in
equations \eqref{DP20} and \eqref{DP21}. 
Here we consider $\tau$ as dimensionless variable 
in order to have standard physical dimensions of physical quantities. 
Equations \eqref{DP20} and \eqref{DP21} are fractional Heisenberg equations \cite{M8} for the Dirac particle.


We should note that we use the power-law memory function 
since it can be considered as an approximation of the equations 
with generalized memory functions.
In the paper \cite{GM}, using the generalized Taylor series in the Trujillo-Rivero-Bonilla form 
for the memory function, we proved that the equations with memory functions 
can be represented through the Riemann-Liouville fractional integrals 
(and the Caputo fractional derivatives) of non-integer orders
for wide class of the memory functions.

Note that the proposed generalized Heisenberg equations \eqref{DP20} and \eqref{DP21} with $\alpha>1$ can be obtained from the integral equations
\begin{equation} \label{DP23}
\frac{dx_{\mu} (\tau)}{d \tau} = 
- 2 \int^{\tau}_0 M_I(\tau - \tau_1) 
\pi_{\mu} (\tau_1) d \tau_1 ,
\end{equation}
\begin{equation} \label{DP24}
\frac{d\pi_{\mu} (\tau)}{d \tau} = 
 \int^{\tau}_0 M_I(\tau - \tau_1) 
\left( - 2 eF_{\mu \nu} \pi^{\nu} (\tau_1) -
ie \partial^{\nu} F_{\mu \nu}
- \frac{e}{2} \partial_{\mu} F_{\nu \rho} \sigma^{\nu \rho} \right) d \tau_1 ,
\end{equation}
where
\begin{equation} \label{DP25}
M_I (\tau - \tau_1) =\frac{1}{\Gamma (\beta)} (\tau - \tau_1)^{\beta-1} .
\end{equation}
This possibility is based on the fact that the Caputo fractional derivative is the left inverse operator for the Riemann-Liouville fractional integral (see equation of the Lemma 2.21 of \cite{FC4}, p.95), the has the form
\begin{equation} \label{DP26} 
\left(D^{\beta}_{\tau,0+} \left(I^{\beta}_{\tau,0+} {\cal A}_{\mu} \right)\right)(\tau) = {\cal A}_{\mu}(\tau) ,
\end{equation}
if $\beta>0$, and ${\cal A}_{\mu}(\tau) \in L_{\infty}(0,T)$ or ${\cal A}_{\mu}(\tau) \in C[0,T]$.

The action of the Caputo derivative on equations \eqref{DP23} and \eqref{DP24} gives the suggested equations \eqref{DP20} and \eqref{DP21} with the order $\alpha=\beta+1$. 

As a result, we can formulate the following statement.

{\bf Statement 1.} \\
{\it
(a) For Dirac particle in external electromagnetic field, the memory effects (the proper time nonlocality) can be described by equations \eqref{DP15} and \eqref{DP16}, or equations \eqref{DP23} and \eqref{DP24}. \\
(b) The Dirac particle with memory, which is described by
the power-law memory function \eqref{DP18}, 
is described by the fractional differential equations
\eqref{DP20} and \eqref{DP21} of the order $\alpha>0$. \\
(c) The standard equations \eqref{DP12} and
\eqref{DP13} of the Dirac particle without memory
are special cases of equations \eqref{DP20} and \eqref{DP21} for  $\alpha=1$.
}

Note that the proper time nonlocality and memory effects can be caused by the interaction of the particle with environment \cite{CEJP2012}.


\section{Violation of semi-group property: Non-Markovian dynamics}

In this section, we discuss the characteristic property of dynamics with memory.

Let us consider the fractional differential equations
\begin{equation} \label{CF-11new} 
\left(D^{\alpha}_{\tau;0+} {\cal A}^{\mu} \right)(\tau) = 
{\Lambda^{\mu}}_{\nu} \, {\cal A}^{\nu}(\tau) ,
\end{equation} 
where $0<\alpha \le 1$. 
For example, we can consider 
${\cal A}(\tau)=\{\pi^{\mu}(\tau)\}$ and
${\Lambda_{\mu}}_{\nu} = - 2 e {F^{\mu}}_{\nu}$.
Using the matrix notation 
$\Lambda=\{{\Lambda^{\mu}}_{\nu}\}$ and
${\cal A}(\tau)=\{A^{\mu}(\tau)\}$, 
equation \eqref{CF-11new} 
can be represented as the matrix equation
\begin{equation} \label{CF-11new2} 
\left(D^{\alpha}_{\tau;0+} {\cal A} \right)(\tau) = 
\Lambda \, {\cal A} (\tau) . 
\end{equation} 

The Cauchy-type problem for equation \eqref{CF-11new2}, 
in which the initial condition is given 
at the time $\tau=0$ by ${\cal A}(0)$,
has solution that can be represented in the form
\begin{equation} \label{ML-1} 
{\cal A} (\tau) = \Phi_{\tau} (\alpha) {\cal A} (0) , 
\quad (\tau \ge 0), 
\end{equation} 
where 
\begin{equation} \label{ML-2}
\Phi_{\tau}(\alpha) = 
E_{\alpha} \left[ \Lambda \tau^{\alpha} \right] . 
\end{equation}
Here $E_{\alpha} \left[ \Lambda \tau^{\alpha} \right]$ 
is the Mittag-Leffler function \cite{MLF} with the matrix argument
\begin{equation} \label{ML-3} 
E_{\alpha} \left[ \Lambda \tau^{\alpha} \right] = 
\sum^{\infty}_{k=0} \frac{\tau^{\alpha k}}{\Gamma(\alpha k +1)} 
\Lambda^k . 
\end{equation}

The operator $\Phi_{\tau}(\alpha)$ describes dynamics of open quantum systems with power-law memory.
The matrix $\Lambda$ can be considered as a 
generator of the one-parameter groupoid $\Phi_{\tau}(\alpha)$
on operator algebra of quantum observables
\begin{equation} \label{ML-4} 
\left( D^{\alpha}_{\tau,0+} \Phi_{\tau}(\alpha) \right)(\tau) =
\Lambda \Phi_{\tau}(\alpha) . 
\end{equation}
The set $\{ \Phi_{\tau}(\alpha) | \, \tau \geqslant 0\}$, 
can be called a quantum dynamical groupoid \cite{AP2012}.
Note that the following properties are realized
\begin{equation} \label{ML-5} 
\Phi_{\tau}(\alpha) I=I , \quad
(\Phi_{\tau}(\alpha) {\cal A})^*= \Phi_{\tau}(\alpha) {\cal A}
\end{equation}
for self-adjoint operators 
${\cal A}$ (${\cal A}^*={\cal A}$), and
\begin{equation} \label{ML-7} 
\lim_{\tau \to 0+} \Phi_{\tau}(\alpha) = I , 
\end{equation}
where $I$ is an identity operator ($I {\cal A}={\cal A}$).
As a result, the operators $\Phi_{\tau}(\alpha)$, $\tau \geqslant 0$, 
are real and unit preserving maps on operator algebra of quantum observables.

For $\alpha=1$, we have
\begin{equation} \label{ML-8} 
\Phi_{\tau}(1)= E_{1} [ \Lambda \tau] = 
\exp \{ \Lambda \tau \} . 
\end{equation} 
The operators $\Phi_{\tau}=\Phi_{\tau}(1)$ form a semigroup such that
\begin{equation} \label{ML-9} 
\Phi_{\tau} \Phi_{s}=\Phi_{\tau+s}, \quad 
(\tau,s >0) , \quad \Phi_0=I . 
\end{equation} 
This property holds since
\begin{equation} \label{ML-10} 
\exp \{ \Lambda \tau \} \ \exp \{ \Lambda s \} =
\exp \{ \Lambda (\tau+s) \} . 
\end{equation}

For $\alpha \not \in \mathbb{N}$ we have
\begin{equation} \label{ML-11} 
E_{\alpha} [\Lambda \tau^{\alpha} ] \,
E_{\alpha} [ \Lambda s^{\alpha} ] \not=
E_{\alpha} [ \Lambda (\tau+s)^{\alpha} ] . 
\end{equation}
Therefore the semigroup property \cite{MLF1,MLF2,MLF3}
is not satisfied for non-integer values of $\alpha$:
\begin{equation} \label{ML-12} 
\Phi_{\tau} (\alpha) \Phi_s (\alpha) \not=
\Phi_{\tau+s} (\alpha), \quad (\tau,s >0) . \end{equation} 

As a result, we can formulate the following statement.

{\bf Statement 2.} \\
{\it The operators $\Phi_{\tau}(\alpha)$,
which describe quantum dynamics of Dirac particle with memory,
cannot form a semi-group in general, 
since the semigroup property is violated for $\alpha \not \in \mathbb{N}$.
}

This property means that equations \eqref{DP20} and \eqref{DP21} for Dirac particle with memory describe non-Markovian evolution.
Note that the violation of this semigroup property is 
a characteristic property of dynamics with memory.


\section{Dynamics with memory in constant electromagnetic field}

\subsection{Equations for constant electromagnetic field}


Let us consider a constant electromagnetic field. 
The equations, which described Dirac particle with power-law memory in a constant field, can be written as 
\begin{equation} \label{CF-1} 
\left( {\cal D}^{\alpha}_{\tau,0+} 
x^{\mu} (\tau) \right) (t) = -2 \pi^{\mu} (\tau) ,
\end{equation} 
\begin{equation} \label{CF-2}
\left( {\cal D}^{\alpha}_{\tau,0+} 
\pi^{\mu} (\tau) \right) (t) = - 2 e {F^{\mu}}_{\nu} \pi^{\nu} (\tau) .
\end{equation} 

For $\alpha=1$ equations \eqref{CF-1} and \eqref{CF-2} are reduced to 
\begin{equation} \label{CF-3}
\frac{dx^{\mu} (\tau)}{d \tau} = - 2 \pi^{\mu} (\tau) ,
\end{equation}
\begin{equation} \label{CF-4}
\frac{d\pi^{\mu} (\tau)}{d \tau} = 
 - 2 e {F^{\mu}}_{\nu} \pi^{\nu} .
\end{equation}
Equations \eqref{CF-3} and \eqref{CF-4} describe the Dirac particle without memory in constant electromagnetic field. 

Using matrix notations $x=\{x^{\mu}\}$, $\pi=\{\pi^{\mu}\}$,
$F=\{{F^{\mu}}_{\nu}\}$, equations \eqref{CF-3} and \eqref{CF-4} can be integrated, and we get the well-known solutions \cite{IZ} in the form
\begin{equation} \label{CF-5}
\pi(\tau) = e^{-2e F \tau} \pi(0) ,
\end{equation}
\begin{equation} \label{CF-6}
x(\tau) = x(0) + 
\left( \frac{e^{-2e F \tau} -1}{eF} \right) \pi(0) .
\end{equation}

Let us obtain solutions of the fractional differential equations \eqref{CF-1} and \eqref{CF-2}.
 

\subsection{Solution of fractional differential equation for $\pi^{\mu}(\tau)$}

Equation \eqref{CF-2} can be rewritten in the form
\begin{equation} \label{CF-2b}
\left( {\cal D}^{\alpha}_{\tau,0+} 
\pi^{\mu} (\tau) \right) (t) = - 2 e {F^{\mu}}_{\nu} \pi^{\nu} ,
\end{equation} 
where $n-1<\alpha \le n$. 
Using ${\cal A}^{\mu}(\tau)=\pi^{\mu}(\tau)$, and
${\Lambda^{\mu}}_{\nu}=- 2 e {F^{\mu}}_{\nu} $,
equation \eqref{CF-2b} is written in the form
\begin{equation} \label{CF-2c} 
\left(D^{\alpha}_{\tau;0+} {\cal A}^{\mu} \right)(\tau) = 
{\Lambda^{\mu}}_{\nu} \, {\cal A}^{\nu}(\tau) .
\end{equation} 
Using the matrix notation 
$\Lambda=\{{\Lambda^{\mu}}_{\nu}\}$ and
${\cal A}(\tau)=\{A^{\mu}(\tau)\}$, 
equation \eqref{CF-2c} can be represented as the matrix equation
\begin{equation} \label{CF-11m} 
\left(D^{\alpha}_{\tau;0+} {\cal A} \right)(\tau) = 
\Lambda \, {\cal A} (\tau) .
\end{equation}

To solve equation \eqref{CF-11m}, we can use Theorem 5.15 of book \cite{FC4}, p.323. 
The solution of equation \eqref{CF-11m} has the form
\begin{equation} \label{CF-12} 
{\cal A} (\tau) = \sum^{n-1}_{k=0} \tau^{k}
 \, E_{\alpha,k+1} \left[ \Lambda \, \tau^{\alpha} \right]
 {\cal A}^{(k)}(0) , 
\end{equation} 
where $n-1<\alpha \le n$, and $E_{\alpha ,\beta}[z]$ is the two-parameter Mittag-Leffler function \cite{FC4,MLF} that is defined by the expression 
\begin{equation} \label{CF-13} 
E_{\alpha ,\beta} 
\left[ \Lambda \, \tau^{\alpha} \right] = 
\sum^{\infty}_{k=0} 
\frac{\tau^{k\alpha}}{\Gamma (\alpha k + \beta)} \Lambda^k , 
\end{equation} 
where $\Gamma(\alpha)$ is the gamma function, 
$\alpha >0$, and $\beta$ is an arbitrary real number.
For $0<\alpha <1$, solution \eqref{CF-12} takes the form 
\begin{equation} \label{CF-14}
{\cal A}(\tau) = 
E_{\alpha,1} \left[\Lambda \tau^{\alpha}\right] 
{\cal A}(0) . 
\end{equation}
For $\alpha=1$, we can use $E_{1,1}[z]=e^{z}$. Then 
solution \eqref{CF-12} gives the equation 
\begin{equation} \label{CF-15} 
{\cal A} (\tau) = \exp \{ \Lambda \tau \} 
{\cal A} (0) .
\end{equation}

As a result, we can formulate the following statement.

{\bf Statement 3.} \\
{\it In the case of constant external electromagnetic field ($F=const$), 
solution of fractional differential equation \eqref{CF-2} (or \eqref{CF-2b}) for the operator $\pi^{\mu}(\tau)$ has the form 
\begin{equation} \label{CF-16} 
\pi(\tau) = 
\sum^{n-1}_{k=0} \tau^{k}
 \, E_{\alpha,k+1} \left[ - 2 e F \, \tau^{\alpha} \right] 
\pi^{(k)}(0) ,
\end{equation}
where $F=\{ {F^{\mu}}_{\nu}\}$ and $n-1<\alpha \le n$. 
}

Expression \eqref{CF-16} describes dynamics of Dirac particle with memory in constant electromagnetic field.

For $0<\alpha \le 1$, equation \eqref{CF-16} takes the form
\begin{equation} \label{CF-16b} 
\pi(\tau) = E_{\alpha,1} \left[ - 2 e F \tau^{\alpha} \right]
\pi(0) . 
\end{equation}

For $\alpha=1$, solution \eqref{CF-16} takes the standard form \eqref{CF-5}, i.e., $\pi(\tau) = e^{-2e F \tau} \pi(0)$.


\subsection{Eigenvalues of electromagnetic tensor for constant fields}

Let us consider the equation \eqref{CF-16} that has the form
\begin{equation} \label{CF-16c} 
\pi(\tau) = 
\sum^{n-1}_{k=0} \tau^{k}
 \, E_{\alpha,k+1} \left[ \Lambda \, \tau^{\alpha} \right] 
\pi^{(k)}(0) ,
\end{equation}
where $\Lambda$ is the matrix 
\begin{equation} \label{CF-16c2} 
\Lambda = \{ -2 e {F^{\mu}}_{\nu} \}= - 2 e 
\left( \begin{array}{cccc}
0 & E^1 & E^2 & E^3 \\ 
E^1 & 0 & B^3 & -B^2 \\ 
E^2 & - B^3 & 0 & B^1 \\ 
E^3 & B^2 & -B^1 & 0 
\end{array}
\right) .
\end{equation}
To get expression \eqref{CF-16c2}, we use that 
${F^{\mu}}_{\nu}=F^{\mu \rho} g_{\rho \nu}$ and
\begin{equation} \label{CF-16g} 
\{ g^{\mu \nu}\} = \{ g_{\mu \nu} \}= 
\left( \begin{array}{cccc}
1 & 0 & 0 & 0 \\ 
0 & -1 & 0 & 0 \\ 
0 & 0 & -1 & 0 \\ 
0 & 0 & 0 & -1 
\end{array}
\right) ,
\end{equation}
that gives the equations
\begin{equation} \label{CF-16FF}
{F^{\mu}}_0 = F^{\mu 0} , \quad 
{F^{\mu}}_k = - F^{\mu k} \quad (k=1,2,3; \quad \mu=0,1,2,3) .
\end{equation}


We can use the diagonalization of the real skew-symmetric matrix $\Lambda^{\mu}_{\nu} = 2 e F^{\mu}_{\nu}$, where $F^{\mu}_{\nu}=g^{\mu \rho} F_{\rho \nu}$, if all eigenvalues of this matrix are nonzero. 
Note that the $4 \times 4$ matrix 
$\Lambda=\{\Lambda^{\mu}_{\nu}\}$ is not skew-symmetric matrix. 

An $n\times n$ matrix $\Lambda$ over a field $R$ is diagonalizable if and only if there exists a basis of $R^n$ consisting of eigenvectors of $ \Lambda$. If such a basis has been found, we can form the matrix $N$ having these basis vectors as columns.
Then there exists an invertible matrix $N$, which is called a modal matrix for $\Lambda$, such that $L=N\Lambda N^{-1}$ is a diagonal matrix. 
The diagonal elements $\lambda_k$, $k=1,2,...,n$, of this matrix $L = diagonal \{\lambda_0,\lambda_1,\lambda_2,\lambda_3\}$ 
are the eigenvalues of $\Lambda$.
An $n\times n$ matrix $\Lambda$ is diagonalizable if it has $n$ distinct eigenvalues, i.e. if its characteristic polynomial has $n$ distinct roots. 

It is known that the eigenvalues of the matrix 
$F={F^{\mu}}_{\nu}= F^{\mu \rho} g_{\rho\nu}$, 
are the roots of the fourth-degree polynomial 
with invariants $(1/2) F_{\mu \nu} \, ^*F^{\mu \nu}=(\vec{E},\vec {B})$ and $(1/2)F_{\mu \nu} F^{\mu \nu} = 
\left( B^2-E^2 \right)$ in the coefficients (for example see \cite{Fradkin,Yaremko}).
Therefore, the eigenvalues of the matrix $\Lambda =-2eF$
are the roots of the equation 
\begin{equation} \label{Pol} 
\lambda^4+S \lambda^2- \frac{1}{4} P^2 =0 ,
\end{equation}
where $S$ and $P$ are the invariants 
\begin{equation} \label{SP} 
P = 2 e^2 F_{\mu \nu} \, ^*F^{\mu \nu} 
= 2 e^2 (\vec{E},\vec {B}) , 
\quad 
S = 2 e^2 F_{\mu \nu} F^{\mu \nu} = 
2 e^2 \left( B^2-E^2 \right) , 
\end{equation}

The set of eigenvalues $\lambda_{s}$, where $s=0,1,2,3$
consists of two pairs, one real and the other imaginary
\begin{equation}
\lambda_{s} \in \{ +b,-b; +i a , -i a \} ,
\end{equation}
where $a$ and $b$ are the scalar functions of invariants
\eqref{SP} in the form
\begin{equation} \label{EV=a}
a=\sqrt{ \frac{1}{2} \left( S+\sqrt{S^2+P^2} \right) } ,
\end{equation}
\begin{equation} \label{EV=b}
b=\sqrt{\frac{1}{2} \left(-S+\sqrt{S^2+P^2} \right)} .
\end{equation}
For details see \cite{Fradkin,Yaremko}.


It should be noted that the matrix $\Lambda$ can be represented in the form
\begin{equation} \label{CF-16L1} 
\Lambda = N^{-1} \, L \, N , 
\end{equation} 
where
\begin{equation} \label{CF-16N} 
N = columns \{ \xi_0 ,\xi_1 ,\xi_2 ,\xi_3 \} , \quad
\end{equation} 
\begin{equation} \label{CF-16L2} 
L = diagonal \{\lambda_0,\lambda_1,\lambda_2,\lambda_3 \} = 
\left( \begin{array}{cccc}
\lambda_0 & 0 & 0 & 0 \\ 
0 & \lambda_1 & 0 & 0 \\ 
0 & 0 & \lambda_2 & 0 \\ 
0 & 0 & 0 & \lambda_3 
\end{array}
\right), 
\end{equation} 
and $\lambda_{\mu}$ are the complex coefficients such that
\begin{equation} \label{CF-16L3} 
\Lambda \xi_{s} = \lambda_{s} \xi_{s} .
\end{equation} 
Here $\xi_{s}$, $s=0,1,2,3$, are eigenvectors and $\lambda_{s}$
are eigenvalues of the matrix $\Lambda$, respectively.
In equation \eqref{CF-16L3} there is no sum on the index $\rho$.

As a result, we can formulate the following statement.

{\bf Statement 4.} \\
{\it In the case of constant external electromagnetic field ($F=const$), which does not change with proper time, the solution \eqref{CF-16c} of fractional differential equation \eqref{CF-2} (or \eqref{CF-2b}) of the order $n-1<\alpha \le n$ for the operator $\pi^{\mu}(\tau)$ can be written in the form
\begin{equation} \label{CF-16c5} 
\pi(\tau) = \sum^{n-1}_{k=0} \tau^{k} N^{-1} \, 
E_{\alpha,k+1} \left[ L \, \tau^{\alpha} \right] \, N 
\pi^{(k)}(0) .
\end{equation}
where $E_{\alpha ,\beta}[z]$ is the two-parametric Mittag-Leffler function \eqref{ML-3}. 
}

Equation \eqref{CF-16c5} describes the evolution of the operator $\pi^{\mu}(\tau)$ of the Dirac particle with power-law memory that is considered as an open quantum system.


\subsection{Asymptotic behavior of $\pi^{\mu}(\tau)$}

Let us describe the asymptotic behavior of solution \eqref{CF-16} 
at $\tau \to \infty$. 
To describe the behavior of the two-parameter Mittag-Leffler function 
$E_{\alpha ,k+1}\left[\lambda \, \tau^{\alpha}\right]$ at 
$\tau \to \infty $, we can use 
equations 1.8.27 and 1.8.28 of book \cite{FC4}, p.43.

For $|arg(\lambda)|\le m_0$ and $0<\alpha <2$, where
\[
\frac{\pi \alpha}{2} < m_0 < min\{ \pi, \pi \alpha\} ,
\]
we have the asymptotic behavior 
of the two-parameter Mittag-Leffler function in the form
\begin{equation} \label{CF-17} 
E_{\alpha , \beta +1} 
\left[\lambda \, \tau^{\alpha}\right] = 
\frac{\lambda^{- \beta/\alpha}}{\alpha}{ \, \tau^{-\beta}
 \, \exp \left( \lambda^{1/\alpha} \, \tau\right) \ } - 
\sum^{m}_{j=1} \frac{\lambda^{-j}}{
\Gamma (\beta +1-\alpha \, j)} \, 
\frac{1}{\tau^{\alpha \, j}}
+O \left(\frac{1}{\tau^{\alpha \, (m+1)}}\right) .
\end{equation} 
For $m_0<|arg(\lambda)|\le \pi$ and $0<\alpha <2$, we have the asymptotic behavior in the form
\begin{equation} \label{CF-17b} 
E_{\alpha , \beta +1} 
\left[\lambda \, \tau^{\alpha}\right] = 
- \sum^{m}_{j=1} \frac{\lambda^{-j}}{
\Gamma (\beta +1-\alpha \, j)} \, 
\frac{1}{\tau^{\alpha \, j}}
+O \left(\frac{1}{\tau^{\alpha \, (m+1)}}\right) .
\end{equation}

Therefore we can use the following equations for the asymptotic behavior of the two-parameter Mittag-Leffler functions, which are part of the solutions.

For the eigenvalue $\lambda_s=b>0$, where $b$ is defined by \eqref{EV=b}
(i.e. the case $\lambda=Re[\lambda]>0$ with $Im[\lambda]=0$), and $0<\alpha <2$, we should use equation
\begin{equation} \label{CF-17c} 
E_{\alpha , \beta +1} 
\left[\lambda \, \tau^{\alpha}\right] = 
\frac{\lambda^{- \beta/\alpha}}{\alpha} \, \tau^{-\beta}
 \, \exp \left( \lambda^{1/\alpha}_s \, \tau\right) - 
\sum^{m}_{j=1} \frac{\lambda^{-j}}{
\Gamma (\beta +1-\alpha \, j)} \, 
\frac{1}{\tau^{\alpha \, j}}
+O \left(\frac{1}{\tau^{\alpha \, (m+1)}}\right) 
\end{equation} 
for $\tau \to \infty $, where $\lambda$ is a positive real number.

For the eigenvalues $\lambda_s=\pm i a$ and $\lambda_s=-b<0$, where $a$ and $b$ are defined by \eqref{EV=a}, \eqref{EV=b}
(i.e. the case $\lambda=Re[\lambda]=0$ with $Im[\lambda] \ne 0$, and the case $\lambda=Re[\lambda]<0$ with $Im[\lambda]=0$), 
and $0<\alpha <2$, we should use equation
\begin{equation} \label{CF-17d} 
E_{\alpha , \beta +1} 
\left[\lambda \, \tau^{\alpha}\right] = 
- \sum^{m}_{j=1} \frac{\lambda^{-j}_s}{
\Gamma (\beta +1-\alpha \, j)} \, 
\frac{1}{\tau^{\alpha \, j}}
+O \left(\frac{1}{\tau^{\alpha \, (m+1)}}\right) 
\end{equation}
for $\tau \to \infty $, where $\lambda$ is a negative real number or $\lambda$ is an imaginary number.
The difference between the formulas \eqref{CF-17b} and
\eqref{CF-17} is the absence of the first term with the exponent. 


Asymptotic equations \eqref{CF-17c} and \eqref{CF-17d} allow us to describe the behavior at $\tau \to \infty $ in model with power-law memory fading parameter $0<\alpha<1$ and $1<\alpha<2$. Substitution of expressions \eqref{CF-17c} and \eqref{CF-17d} with $\beta =k$ into \eqref{CF-16c} can give the asymptotic expressions of solution.

As a result, we can formulate the following statement.

{\bf Statement 5.} \\
{\it 
The asymptotic behavior of the variable
\begin{equation} \label{CF-18P}
\Pi_s (\tau) = (N \pi)_s(\tau) = {N^s}_{\mu} \pi^{\mu} (\tau) ,
\end{equation} 
which characterizes the Dirac particle with memory in a constant field, is described by the following equations.
For the eigenvalues $\lambda_s=\pm i a$ and $\lambda_s=-b<0$, where $a$ and $b$ are defined by \eqref{EV=a}, \eqref{EV=b}
and $0<n-1 <\alpha <n$, we have the equation
\begin{equation} \label{CF-18} 
\Pi_{s} (\tau) = - \sum^{n-1}_{k=0} 
\left( \sum^{m}_{j=1} 
\frac{ \tau^{k-\alpha \, j} \lambda^{-j}_s }{\Gamma (k+1-\alpha \, j)} \, \Pi^{(k)}_s (0) 
 +
O \left( \frac{1}{\tau^{\alpha \, (m+1) - k}}\right) \right) . 
\end{equation} 
For the eigenvalue $\lambda_s=b>0$, where $b$ is defined by \eqref{EV=b} and $0<n-1 <\alpha <n$, 
the asymptotic behavior of the variable \eqref{CF-18P} 
is described by the equation 
\begin{equation} \label{CF-18b} 
\Pi_{s} (\tau) = \sum^{n-1}_{k=0} 
\left( \frac{\lambda^{- k/\alpha}}{\alpha} 
 \, \exp \left( \lambda^{1/\alpha}_s \, \tau\right) -
\sum^{m}_{j=1} 
\frac{ \tau^{k-\alpha \, j} \lambda^{-j}_s }{\Gamma (k+1-\alpha \, j)} \, \Pi^{(k)}_s (0) 
 +
O \left( \frac{1}{\tau^{\alpha \, (m+1) - k}}\right) \right) . 
\end{equation} 
}


Expressions \eqref{CF-18} and \eqref{CF-18b} describe the behavior of $\pi(\tau)$ at $\tau \to \infty $ for the Dirac particle with power-law memory in constant external electromagnetic field.


\subsection{Solution of equation for $x^{\mu}(\tau)$}

To solve equation \eqref{CF-1}, we can use Theorem 5.15 of book \cite{FC4}, p.323. Fractional differential equation \eqref{CF-1} can be written in the form 
\begin{equation} \label{CF-7} 
\left(D^{\alpha}_{\tau,0} {\cal A}^{\mu} \right)(\tau) = 
B^{\mu}(\tau) , 
\end{equation} 
where 
${\cal A}^{\mu}(\tau)=x^{\mu}(\tau)$, 
$B^{\mu}(\tau)=-2\pi^{\mu}(\tau)$ and $n-1<\alpha \le n $. 
If $B^{\mu}(\tau)$ are continuous functions defined on the positive semiaxis ($\tau>0$), 
then equations \eqref{CF-7} has the solution 
(for details see Lemma 2.22 and equation 2.4.42 in \cite{FC4}, p.96) in the form
\begin{equation} \label{CF-8} 
{\cal A}^{\mu}(\tau) = \sum^{n-1}_{k=0} \frac{\tau^k}{k!} {{\cal A}^{\mu}}^{(k)} (0)+ \frac{1}{\Gamma (\alpha)}
\int^t_0 (\tau - \tau_1)^{\alpha - 1}
\, B^{\mu}(\tau_1) d \tau_1 . 
\end{equation} 

As a result, we get the equation
\begin{equation} \label{CF-9} 
x^{\mu}(\tau) = \sum^{n-1}_{k=0} \frac{\tau^k}{k!} {x^{\mu}}^{(k)}(0) - \frac{2e}{\Gamma (\alpha)}
\int^t_0 (\tau - \tau_1)^{\alpha - 1}
\, \pi^{\mu}(\tau_1) d \tau_1 . 
\end{equation} 
For $0<\alpha<1$, we have
\begin{equation} \label{CF-10} 
x^{\mu}(\tau) - x^{\mu}(0) = \frac{-2e}{\Gamma (\alpha)} 
\int^{\tau}_0 (\tau - \tau_1)^{\alpha - 1}
\, \pi^{\mu}(\tau_1) d \tau_1 . 
\end{equation} 
Note that equation \eqref{CF-10} can be written through the Riemann-Liouville fractional integral (the definition of this operator is given in \cite{FC4}, p.69-70)


Using the matrix notation $x(\tau)=\{x^{\mu}(\tau)\}$, and
$\pi(\tau)=\{\pi^{\mu}(\tau)\}$ equation takes the form 
\begin{equation} \label{CF-9b} 
x(\tau) = \sum^{n-1}_{k=0} \frac{\tau^k}{k!} {x}^{(k)}(0) - \frac{2e}{\Gamma (\alpha)}
\int^t_0 (\tau - \tau_1)^{\alpha - 1} \, \pi(\tau_1) d \tau_1 . 
\end{equation} 
Substitution of the expression \eqref{CF-16} 
of the solution for $\pi(\tau)$ 
into expression \eqref{CF-9b} of $x(\tau)$ gives
the equation 
\begin{equation} \label{CF-22} 
x(\tau) = 
\sum^{n-1}_{k=0} \frac{\tau^k}{k!} x^{(k)}(0) -
\sum^{n-1}_{k=0} \frac{2e}{\Gamma (\alpha)} 
\int^{\tau}_0 (\tau - \tau_1)^{\alpha - 1} \, \tau^{k}_1 \, 
E_{\alpha,k+1} \left[ \Lambda \, \tau^{\alpha}_1 \right]
\pi^{(k)}(0) d \tau_1 . 
\end{equation} 
where $\Lambda= -2eF =\{ -2e{F^{\mu}}_{\nu}\}$.

To calculate integral of \eqref{CF-22}, we can use equation 4.10.8 from \cite{MLF}, p.86, or equation 4.4.5 of \cite{MLF}, p.61, where $\Gamma (\mu)$ should be used instead of $\Gamma (\alpha)$, in the form
\begin{equation} \label{CF-23} 
\int^{\tau}_0 \tau^{\beta - 1}_1 \, E_{\alpha ,\beta}
\left[\lambda \, \tau^{\alpha}_1\right]
 \, \left(\tau-\tau_1 \right)^{\mu - 1} d \tau_1 = 
\Gamma (\mu) \, \tau^{\beta +\mu - 1} \, 
E_{\alpha ,\beta +\mu} \left[\lambda \, \tau^{\alpha}\right], 
\end{equation} 
where $\beta >0$ and $\mu >0$. 
In equation \eqref{CF-23} we have $\mu=\alpha$, $\beta=k+1$, where the parameter $\mu$ can be considered as a positive integer, or as a positive real number. 


As a result, we can formulate the following statement by using equation \eqref{CF-23} with $\mu=\alpha$ and $\beta=k+1$ for solution \eqref{CF-22}.

{\bf Statement 6.} \\
{\it In the case of constant external electromagnetic field ($F=const$), 
solution of fractional differential equation \eqref{CF-1} of the order $n-1<\alpha \le n$ for the operator $x^{\mu}(\tau)$ has the form 
\begin{equation} \label{CF-24} 
x(\tau) = 
\sum^{n-1}_{k=0} \frac{\tau^k}{k!} x^{(k)}(0) -
2e \sum^{n-1}_{k=0} \tau^{\alpha+k} 
E_{\alpha ,\alpha+k+1} \left[\Lambda \, \tau^{\alpha}\right] 
\pi^{(k)} (0) ,
\end{equation} 
where $\Lambda=-2eF$.
}

Note that using equation \eqref{CF-23}, we can easily generalize solutions to the case of different memory fading parameters for variables $x^{\mu}(\tau)$ and $\pi^{\nu}(\tau)$ (for example, when the order of the Caputo fractional derivative in equation \eqref{CF-7} is equal to $\mu$ instead of $\alpha$).

For $0<\alpha \le 1$, expression \eqref{CF-24} takes the form
\begin{equation} \label{CF-25} 
x(\tau) = x(0) - 2e \tau^{\alpha} 
E_{\alpha ,\alpha+1} \left[\Lambda \, \tau^{\alpha}\right] 
\pi (0) .
\end{equation} 
Using that $\Lambda=-2eF$, equation \eqref{CF-25} 
can be rewritten as
\begin{equation} \label{CF-25a1} 
x(\tau) = x(0) - 2e \tau^{\alpha} 
E_{\alpha ,\alpha+1} \left[-2e F \, \tau^{\alpha}\right] 
\pi (0) ,
\end{equation} 
where $0<\alpha \le 1$.
Using equations 1.8.2 in \cite{FC4}, p.40,
 and 1.8.19 in \cite{FC4}, p.42, 
(see also equation 4.2.1 in \cite{MLF}, p.57) in the form
\begin{equation} \label{MLF2} 
E_{1,1} [z] = E_{1} [z] = e^z , \quad
E_{1,2} [z] = \frac{e^z-1}{z} 
\end{equation} 
the proposed solutions \eqref{CF-25} 
give the standard solutions 
\eqref{CF-6} for the case $\alpha=1$.


Equations \eqref{CF-16c5} and \eqref{CF-24} describe the Dirac particle with power-law memory as an open quantum system.

\subsection{Asymptotic behavior of $x^{\mu}(\tau)$}

The asymptotic behavior of $x_{\mu}(\tau)$, which is described by expression \eqref{CF-24}, can be obtained by using equations \eqref{CF-17} and \eqref{CF-17b} for $\beta = \alpha + k$.

Using the notations
\begin{equation} 
X_s(\tau) = (Nx)_{s} (\tau) = {N^s}_{\mu} x^{\mu} (\tau), \quad 
\Pi_s (\tau) = (N\pi)_{s} (\tau) = {N^s}_{\mu} \pi^{\mu} (\tau) ,
\end{equation} 
we can formulate the following statement.

{\bf Statement 7.} \\
{\it 
For the eigenvalue $\lambda_s=b>0$, where $b$ is defined by \eqref{EV=b} and $0<\alpha <2$,
equations \eqref{CF-17} gives the asymptotic behavior 
of the operator $x^{\mu}(\tau)$ in the form
\begin{equation} \label{AX-1} 
\begin{gathered}
X_s(\tau) = \sum^{n-1}_{k=0} \frac{\tau^k}{k!} X^{(k)}_s(0) -
2e \sum^{n-1}_{k=0} \Bigl(
\frac{\lambda^{- (\alpha+k)/\alpha}_s}{\alpha} 
\, \exp \left( \lambda^{1/\alpha}_s \, \tau\right) - \\
\sum^{m}_{j=1} \frac{\lambda^{-j}_s 
\tau^{k-\alpha \, (j-1)}}{\Gamma (k +1-\alpha \, (j-1))} \, \Pi^{(k)}_s (0)
+ O \left(\frac{1}{\tau^{\alpha \, m -k}}\right) 
\Bigr)
\end{gathered}
\end{equation} 
for $\tau \to \infty $, where $\lambda_s$ is a positive real number.

For the eigenvalues $\lambda_s=\pm i a$ and $\lambda_s=-b<0$, where $a$ and $b$ are defined by \eqref{EV=a}, \eqref{EV=b},
and $0<\alpha <2$, equations \eqref{CF-17b} gives the asymptotic behavior of $x^{\mu}(\tau)$ in the form
\begin{equation} \label{AX-2} 
X_s(\tau) = 
\sum^{n-1}_{k=0} \frac{\tau^k}{k!} X^{(k)}_s(0) -
2e \sum^{n-1}_{k=0} 
\left( - \sum^{m}_{j=1} \frac{\lambda^{-j}_s
\tau^{k-\alpha \, (j-1)}}{\Gamma (k +1-\alpha \, (j-1))} \, 
\Pi^{(k)}_s (0) +O \left(\frac{1}{\tau^{\alpha \, m -k}}\right) 
\right)
\end{equation} 
for $\tau \to \infty$, where $\lambda_s$ is a negative real number or $\lambda$ is an imaginary number.
}

The difference between the formulas \eqref{AX-1} and
\eqref{AX-2} is the absence of the term with the exponent $\exp \left( \lambda^{1/\alpha}_s \, \tau\right)$.

\section{Conclusion}

In this article, a generalization of the theory of Dirac particle in external electromagnetic field is suggested.
In this generalization we take into account an interaction of the Dirac particle with environment.
In the standard approach the interaction of the Dirac particle with environment is not considered.
The interaction with environment (an openness of the system) is described as proper time non-locality by the memory function.
We use the Fock-Schwinger proper time method to have covariant description of dynamics 
and derivatives of non-integer orders to have non-locality in proper time.
In the proposed theory we take into account that the behavior of the Dirac particle at proper time $\tau$ can depend not only at the present proper time $\tau$, but also on the history of changes at $\tau_1<\tau$.
In this case the Dirac particle is considered an open quantum system, dynamics of which can be characterized as non-Markovian evolution.
We demonstrate the violation of the semigroup property of dynamic maps that is a characteristic property of dynamics with memory.
The fractional differential equation, which describes the Dirac particle with memory, and the expression of its exact solution are suggested. The asymptotic behavior of the proposed solutions is described.

Let us briefly describe three possible generalizations of the proposed model.

To describe the Dirac particle in the case of an explicit dependence of the electromagnetic field strength tensor (or the matrix $\Lambda$) on the proper time, it is necessary to use the generalization of the time-ordered product and the ordered exponential proposed in \cite{M6,T5} for the processes with memory.

The proposed model of the Dirac particle with proper time nonlocality (memory) can be generalized to other type of fading memory by using other type fractional derivatives and integrals.

In order to have a consistent description of the interaction of a particle with its environment, it is necessary to take into account the quantum theory of open quantum systems \cite{Davis,Breuer,TBook2008} and fractional generalizations of quantum Markov equations proposed in 
\cite{TQ1,TQ2,TQ3,TQ4,TQ5,AHEP2014,AHEP2018}.



\end{document}